\documentclass{article}
\usepackage{amsmath,graphicx,mlspconf}
\usepackage{url}
\usepackage{algcompatible}
\usepackage{algorithm}
\usepackage{algpseudocode}
\usepackage{amsfonts, amssymb, amsthm}
\usepackage{xcolor}
\usepackage[noadjust]{cite}
\usepackage[hidelinks]{hyperref}
\usepackage{makecell}
\usepackage{multirow}
\usepackage{booktabs}
\usepackage{diagbox}
\usepackage{cuted}
\usepackage{subcaption}
\usepackage{lipsum}
\usepackage[inline]{enumitem}
\usepackage{siunitx}
\DeclareSIUnit\bit{bit}
\DeclareSIUnit\sps{SPS}

\newif\iftrackrivision

\iftrackrivision
\newcommand{\revise}[2]{{\color{red}\sout{#1}}{\color{blue}#2}}
\else
\newcommand{\revise}[2]{{#2}}
\fi

\newcommand{\superscript}[1]{^{\mathrm{#1}}}

\renewcommand{\emph}[1]{\textit{#1}}

\title{ARTIFICIAL ASMR: A CYBER-PSYCHOLOGICAL APPROACH}
\name{%
   Zexin~Fang$^{\star}$%
   \qquad Bin~Han$^{\star}$%
   \qquad C.~Clark~Cao$^{\dagger}$
   \qquad Hans~D. ~Schotten$^{\star\ddagger}$ \thanks{This work is supported in part by the German Federal Ministry of Education and Research within the project Open6GHub (16KISK003K/16KISK004), in part by the European Commission within the Horizon Europe project Hexa-X (101015956), in part by the Network for the Promotion of Young Scientists at RPTU Kaiserslautern-Landau within the project A-SIREN (Individual Research Funding 2022-2), and in part by the Lam Woo Research Fund at Lingnan University (F871223). B. Han (bin.han@rptu.de) is the corresponding author.}%
}
\address{%
   $^{\star}$RPTU Kaiserslautern-Landau, 
   $^{\dagger}$Lingnan University,
   $^{\ddagger}$German Research Center of Artificial Intelligence%
}

\begin{document}

\maketitle
\begin{abstract}
	The popularity of Autonomous Sensory Meridian Response (ASMR) has skyrockted over the past decade, but scientific studies on what exactly triggered ASMR effect remain few and immature, one most commonly acknowledged trigger is that ASMR clips typically provide rich semantic information. 	With our attention caught by the common acoustic patterns in ASMR audios, we investigate the correlation between the cyclic features of audio signals and their effectiveness in triggering ASMR effects. A cyber-psychological approach that combines signal processing, artificial intelligence, and experimental psychology is taken, with which we are able to quantize ASMR-related acoustic features, and therewith synthesize ASMR clips with random cyclic patterns but not delivering identifiably scenarios to the audience, which were proven to be effective in triggering ASMR effects.
\end{abstract}


\section{INTRODUCTION}
Autonomous Sensory Meridian Response (ASMR), a term coined in the 2010s, is widely used to describe an intriguing phenomenon in which specific visual and auditory stimuli trigger tingling sensations accompanied by positive emotions as well as a feeling of deep relaxation~\cite{BD2015autonomous}. With its blooming cultural popularity and growing commercial market~\cite{Harper2019asmr}, ASMR has attracted emerging research interest~\cite{SWR+2021commercializing}, and its cognitive effect has been verified by significant behavioral and neurological evidences~\cite{PBH+2018more, LGR+2018fMRI}. However, to the best of our knowledge, existing work has only identified some semantic elements that trigger the ASMR effect~\cite{BSD2017sensory}, while the acoustic features of auditory triggers remain poorly understood. Though there is a generic claim that low-frequency sounds are widely observed from effectively triggering ASMR audios, it does not capture the repetitive characteristic.

Inspired by a study that reveals the correlation between the trypophobia-triggering effect and cyclostationary features of images~\cite{CW2013fear}, we suspect that the time-frequency and cyclic features of audios may also play an important role in the triggering of ASMR experience. In this paper, we prove this correlation in a cyber-psychological approach that combines various techniques of signal processing, artificial intelligence, and experimental psychology. More specifically, we apply short-time Fourier transform (STFT) and cyclic spectral analysis on recorded ASMR audio clips to extract their acoustic features. \revise{Training generative adversarial networks (GANs) in a feature-targeted manner, we are able to synthesize artificial ASMR audio without a natural sound source. }{Combining generative advertisal networks (GANs) and a post-processing module, we are able to synthesize ASMR clips based on generated cyclic patterns. }We also design a psychological survey to evaluate the effectiveness of both the recorded and synthesized ASMR audios in triggering ASMR effect on humans, and verify the correlation between this effect and the selected acoustic features.
%
\section{AUDIO DATA PROVISIONING}\label{sec:data}
As the object of our study, we obtained off-the-shelf ASMR audios from non-commercial online open sources. We collected four audios recorded from real sounds of different origins, including:
\begin{enumerate*}[label=\emph{\roman*)}]
	\item breathing,
	\item mixing soft cream,
	\item puffing a spray, and
	\item clicking a keyboard.
\end{enumerate*}
Every audio was recorded \SI{16}{\bit} stereo at the sampling rate of \SI{22.05}{\kilo\Hz}, lasting about \SI{1}{\hour}. We intentionally selected these four sound types as our experiment objects because they are largely \revise{emotionally neutral. Indeed, other common types of available ASMR audios, such like whispers, campfires, winds, chewing sounds, ocean waves, etc. usually deliver rich semantic information in their associated imagery, which may cause ASMR-independent emotional effects, and therewith generate biases in the psychological test that we are introducing in Sec.~\ref{subsec:behavior_study}.}{distinct in cyclic and spectral features, for instance, clicking a keyboard clips generally have dense cyclic patterns and more high frequency components, meanwhile breathing clips manifests sparse cyclic patterns with low pass characteristic. }

\section{ANALYSIS AND FEATURE EXTRACTION}\label{sec:audio_analysis}
\begin{figure} [!htpb]
	\centering
	\includegraphics[width=0.45\linewidth]{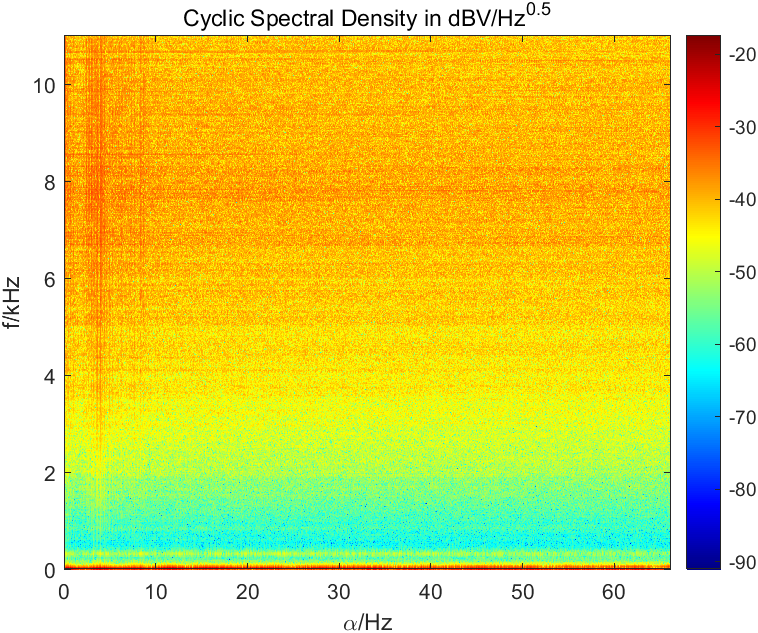}
	\hfill
	\includegraphics[width=0.45\linewidth]{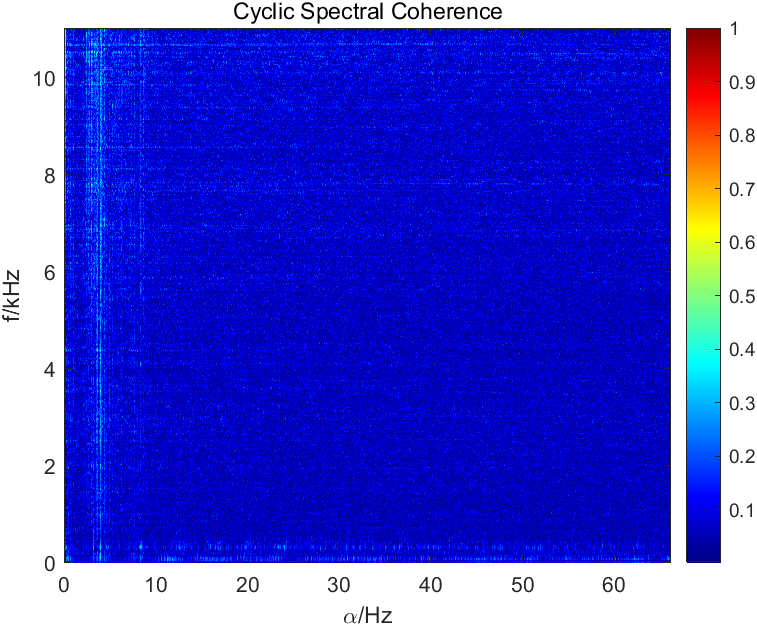}
	\caption{SCD (left) and CCF (right) of the spray audio clip}
	\label{fig:SCD_CCF}
\end{figure}

\revise{Taking a \SI{10}{\second} clip from the puffing spray audio and sliding a \SI{0.1}{\second} raised cosine window over it with 50\% overlap, an example spectrogram as shown in Fig.~\ref{fig:spectrogram} exhibits a general low-pass feature with a cyclic pattern of broadband bursts.}{Spectrograms of the provisioned ASMR clips generally exhibit a low-pass characteristic and cyclic patterns of boradband bursts, as exampled in Fig.~\ref{fig:spectrogram}.}
\revise{Similar features were also observed in other ASMR audios we collected.}{}To further investigate the periodic patterns within the PSD of an audio signal $x$, the spectral correlation density (SCD) and cyclic coherence function (CCF) can be applied as suggested by~\cite{Gardner2006cyclostationarity}:
\begin{figure}[!htpb]
	\centering
	\includegraphics[width=.7\linewidth]{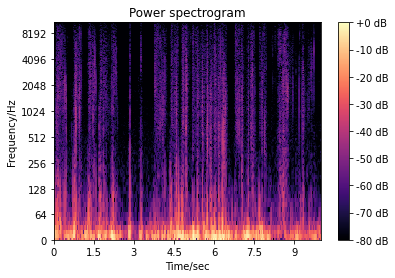}
	\caption{Spectrogram of a puffing spray audio clip}
	\label{fig:spectrogram}
\end{figure}
\begin{align}
	S_{XX}(f,\alpha) &= \mathbb{E}\left\{X(f+\frac{\alpha}{2})X^{*}(f-\frac{\alpha}{2})\right\}\\
	C_{XX}(f,\alpha) &=\frac{S_{XX}(f,\alpha)}{\sqrt{\mathbb{E}\left\{\left\vert X(f+\frac{\alpha}{2})\right\vert^2\right\}\mathbb{E}\left\{\left\vert X(f-\frac{\alpha}{2})\right\vert^2\right\}}}
\end{align}
where $\alpha$ is the cyclic frequency that indicates the periodicity within the spectrum, and $X(f)$ the complex-valued spectrum of $x$ at frequency $f$. For an $N$-sample signal frame $x$, $X$ can be estimated by the Discrete Fourier Transform (DFT):
\begin{equation}
	X(f)=\frac{1}{N}\sum\limits_{k=0}^{N-1}\left[\sum\limits_{n=0}^{N-1}x(n)e^{-\frac{j2\pi nk}{N}}\delta\left(2\pi f-\frac{2\pi k}{N}\right)\right].
\end{equation}
The SCD and CCF of the spray puffing sound clip in Fig.~\ref{fig:spectrogram}, for example, are illustrated in Fig.~\ref{fig:SCD_CCF}. From the figures we see a distinguishing feature that is also observed from other ASMR audios: narrow vertical stripes that distribute smoothly over a wide range in the $f$ domain, while being discrete and sparse in the $\alpha$ domain, occurring at only low cyclic frequencies.

\begin{figure}[!htbp]
	\centering
	\includegraphics[width=0.45\linewidth]{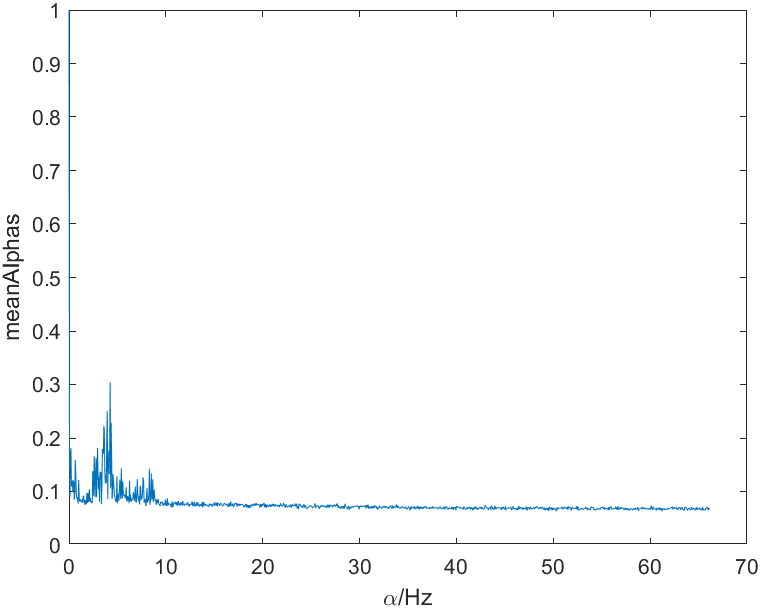}
	\includegraphics[width=0.45\linewidth]{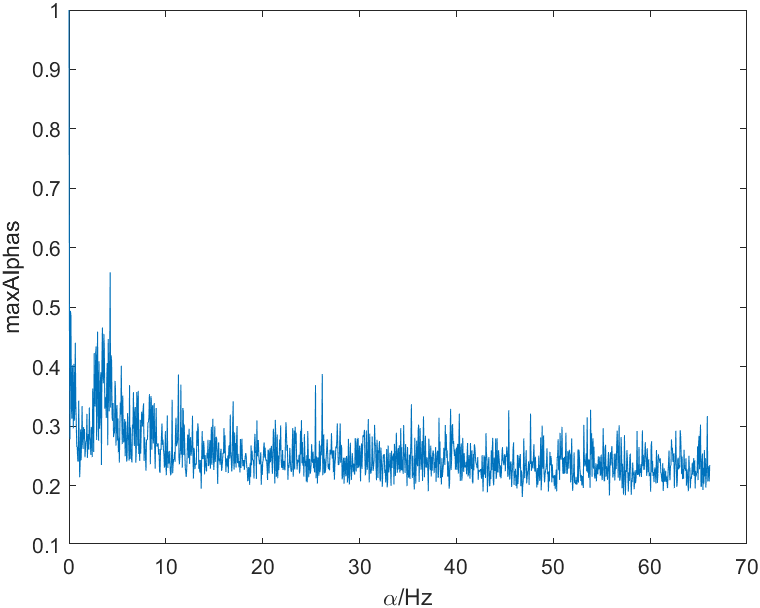}
	\caption{$\overline{S}$ (left) and $S\superscript{max}$ (right) of the spray audio clip}
	\label{fig:alpha_features}
\end{figure}

While both SCD and CCF are bivariate, we further define two univariate functions of $\alpha$ to characterize the cyclic behavior of an audio signal $x$, namely $\overline{S}_{XX}(\alpha)=\frac{1}{N}\sum\limits_{f\in\mathbb{F}}S_{XX}(f,\alpha)$ and
$S\superscript{max}_{XX}(\alpha)=\max\limits_{f\in\mathbb{F}}S_{XX}(f,\alpha)$, where $\mathbb{F}$ is the set of $N$ frequency pins in $S_{XX}$. $\overline{S}_{XX}$ is sensitive to broadband cyclic patterns in $x$, while $S\superscript{max}_{XX}$ is more sensitive to narrow-banded cyclic components, as illustrated in Fig.~\ref{fig:alpha_features}.

On top of $X(f)$, $\overline{S}_{XX}(\alpha)$ and $S\superscript{max}_{XX}(\alpha)$, we define eight features: $\Phi_1(x)={\mathrm{Mean}}\left(\overline{S}_{XX}\right)$, $\Phi_2(x)={\mathrm{Var}}\left(\overline{S}_{XX}\right)$, $\Phi_3(x)={G}\left(\overline{S}_{XX}\right)$, $\Phi_4(x)={\mathrm{Mean}}\left(S\superscript{max}_{XX}\right)$, $\Phi_5(x)={\mathrm{Var}}\left(S\superscript{max}_{XX}\right)$, $\Phi_6(x)={G}\left(S\superscript{max}_{XX}\right)$, $\Phi_7(x)={\mathrm{Var}}\left(\left\vert X\right\vert\right)$, and $\Phi_8(x)={G}\left(\left\vert X\right\vert\right)$, where $G(\mathbb{S})=\frac{\sum\limits_{i\in\mathbb{S}}\sum\limits_{j\in\mathbb{S}}\vert i-j \vert}{2\Vert\mathbb{S}\Vert_0\sum\limits_{i\in\mathbb{S}} i}$ is the Gini coefficient of set $\mathbb{S}$. Here, $\Phi_1$ and $\Phi_4$ assess the overall cyclostationaity, $\Phi_2$ and $\Phi_5$ assess the $\alpha$-variance, $\Phi_3$ and $\Phi_6$ reflect the $\alpha$-sparsity,  $\Phi_7$ describes the $f$-variance of SPD, and $\Phi_8$ the $f$-sparsitiy of SPD.

We sampled three \SI{10}{\second} clips from each recorded audio mentioned in Sec.~\ref{sec:data}, and extracted the features $\Phi_1$--$\Phi_8$ of every clip. A \SI{10}{\second} stereo \SI{22.05}{\kilo\sps} sampled white noise was also analyzed as benchmark. The results are listed in Tab.~\ref{tab:features_recorded}.

\begin{table*}[!htbp]
	\centering
	\caption{Part of features extracted from recorded ASMR audio clips}
	\label{tab:features_recorded}
	\footnotesize
	\begin{tabular}{p{2.5cm}<{\raggedright}p{1.3cm}<{\raggedright}p{1.3cm}<{\raggedright}p{1.3cm}<{\raggedright}p{1.3cm}<{\raggedright}p{1.3cm}<{\raggedright}p{1.3cm}<{\raggedright}p{1.3cm}<{\raggedright}p{1.3cm}<{\raggedright}}
		\Xhline{1.2pt}
		\textbf{Audio Clip} & $\mathbf{\Phi_1}$ & $\mathbf{\Phi_2}$ & $\mathbf{\Phi_3}$ & $\mathbf{\Phi_4}$ & $\mathbf{\Phi_5}$ & $\mathbf{\Phi_6}$ & $\mathbf{\Phi_7}$ & $\mathbf{\Phi_8}$\\ 
		\Xhline{1.2pt}
		White noise & 2,86E-02 & 7,47E-05 & 0,0631 & 8,18E-03 & 3,88E-05 & 0,0426 & 8,20E-03 & 0,5226 \\ 
		Breathing clip 1 & 1,85E-04 & 1,56E-08 & 0,2727 & 1,60E-06 & 6,61E-13 & 0,1190 & 7,51E-04 & 0,9588 \\ 
		Soft cream clip 1 & 4,59E-01 & 1,02E-01 & 0,3262 & 1,13E-03 & 6,40E-07 & 0,2912 & 3,82E-04 & 0,9955 \\  
		Spray clip 1 & 1,20E-03 & 4,81E-07 & 0,2833 & 9,42E-05 & 1,48E-09 & 0,0950 & 1,38E-03 & 0,6393 \\ 
		Keyboard clip 1 & 1,35E-03 & 7,54E-07 & 0,2382 & 3,93E-05 & 1,51E-10 & 0,1091 & 8,20E-04 & 0,9110 \\ 
		\Xhline{1.2pt}
	\end{tabular}
\end{table*}

\section{RANDOM ASMR AUDIO SYNTHESIS}
GAN, since its proposal in 2014~\cite{GPJ+2014generative}, has rapidly demonstrated its great capability of generating artificial data with certain patterns, and therewith attracted intensive attentions from research and development fields of artificial intelligence (AI). In general, a GAN system consists of two neural networks: the \emph{generator} and the \emph{discriminator}. They work against each other and jointly evolve, so that the generator is eventually capable of generating fake data that can be hardly distinguished from real ones~\cite{GPJ+2020generative}. Besides its iconic success in generating fake images, GAN is also proven effective in generating and converting audio signals such as music and speech. Several recent efforts are reported in~\cite{DHY+2018musegan,LKL+2020many,KPL+2022assem,FMW+2022controlling}, which have inspired us to generate random ASMR audios with GAN.


While conventional GAN-based audio solutions are commonly specialized regarding acoustic features of natural languages or musical instruments, they cannot be straightforwardly applied on our problem. Noticing that the spectrogram of an audio signal \begin{enumerate*}[label=\emph{\roman*)}]
\item contains most of the audio information, and
\item exhibits significant 2D patterns upon cyclic audio components,
\end{enumerate*}
we propose to indirectly synthesize ASMR audios by generating artificial ASMR-like \revise{}{cyclic patterns} in spectrograms with GANs. More specifically, concerning the drawbacks of original GANs such like low learning stability and incapability of training with large high-definition datasets, we invoke deep convolutional GANs (DCGANs), which integrate convolutional neural networks (CNNs) into GANs to achieve better stability and efficiency of learning~\cite{RMC15unsupervised,IS2015batch}.

To efficiently train the DCGAN, we set up a training data set with the four recorded ASMR audios introduced in Sec.~\ref{sec:data}. We sliced from every audio $1600$ segments short segments, each with $32~400$ samples that correspond to about $1.5$ seconds, which is sufficient to contain a ASMR-triggering pitch according to our experience. Then we carried out $2048\times2048$ STFT with $50\%$ window overlap, which generated a $1025\times 64$ power spectrogram from every individual segment. To reduce the computational load and stabilize the training, we downsampled every such spectrogram to the size of $192\times 64$, and further divided it into three $64\times 64$ matrices, which is the a data size proven stable to train a 3-channel DCGAN as suggested in~\cite{GSD2020realistic}. This row-wide downsampling was carried out in a bootstrap manner to generate 30 random $64\times 64\times 3$ training samples out of each original power spectrogram. Thus, we obtained $192~000$ training samples in total. \revise{}{To keep the generated cyclic patterns diverse we} clustered the training data into $12$ sets, each mixing samples from different audios \revise{}{of different proportion, for instance set 1 includes more breathing ASMR samples than other kinds meanwhile set 12 has more taping a keyboard ASMR samples}.
In addition, compared to common images where the data values are constrained within a limited quantization range, power spectrogram data usually have a significantly higher dynamic range. To keep details at the lower end of power density, the training samples are log-normalized before linearly rescaled into the range of $[-1,1]$.

To implement the DCGAN, we created a generator model with $4$ deconvolution layers, each applied with a batch normalization, a $ReLU$ activation, a $2\times 2$ stride, and a $1\times 1$ padding. Correspondingly, we created a discriminator model with $5$ convolution layers, all with the same kernel, stride size, and padding size as the generator layers have. We bound both models, initialized their weights w.r.t. a normal distribution $\mathcal{N}(0, 0.02^2)$, and applied an Adam optimizer~\cite{KB2014adam} with learning rate $2\times 10^{-4}$ and $\beta_1=0.5$, $\beta_2=0.999$.

\begin{figure}[!htbp]
\centering
\begin{subfigure}{\linewidth}
	\centering
	\includegraphics[width=.7\linewidth]{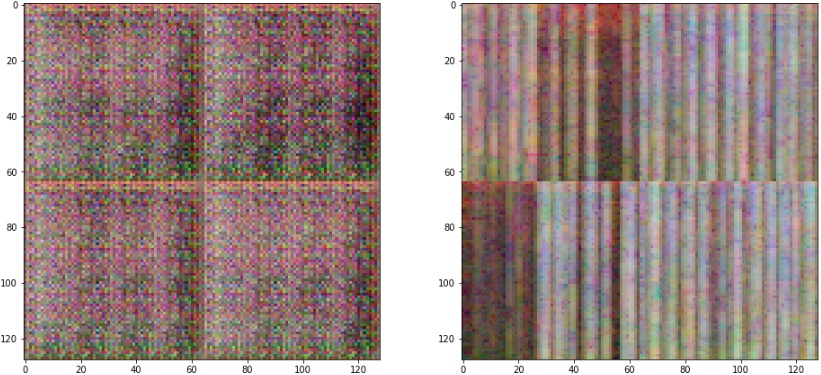}
	\caption{50 iterations}
	\label{subfig:50_iterations}
\end{subfigure}\\
\begin{subfigure}{\linewidth}
	\centering
	\includegraphics[width=.7\linewidth]{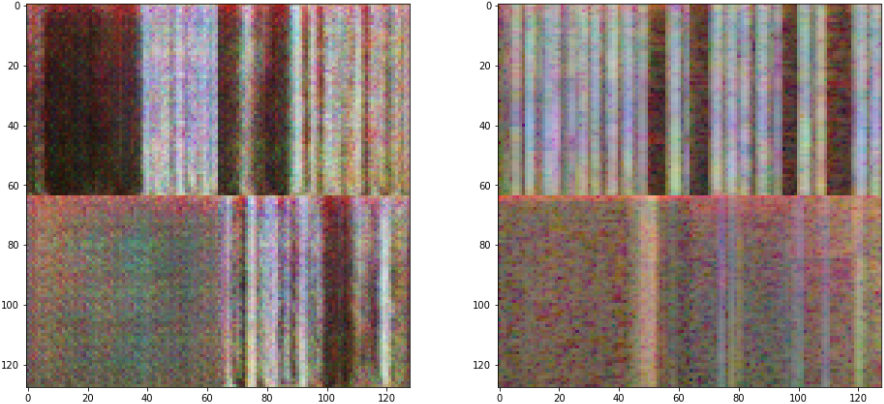}
	\caption{1300 iterations}
	\label{subfig:1300_iterations}
\end{subfigure}	
\caption{Convergence of the DCGAN output: 4 synthesized (left) and 4 recorded (right) spectrogram samples}
\label{fig:training_result}
\end{figure}

After initialization, \revise{we fed the generator network with Gaussian white noise to synthesize artificial samples, which were then sent to the discriminator network together with the recorded samples from a certain training sample group, with a batch size of $64$. The score between $[0,1]$ given by the discriminator is then used to update the generator coefficients in the next iteration. We executed this procedure with all $12$ training sample groups.}{we trained the networks with $12$ sets for $6$ epochs,}
i.e., $1500$ iterations for each set. The network and its output was generally converging through the training process, as exemplified in Fig.~\ref{fig:training_result}.
Upon completion of the training, a $64$-batch of artificial samples was generated at the output of the generator network, each sample representing the spectrogram of a \SI{1.5}{\second} audio segment. From each batch, we randomly selected two such samples and repeatedly concatenated them in a random order to generate spectrograms of longer time frames. In this approach, we obtained $12$ random synthesized \SI{10}{\second} spectrograms.
To reconstruct audio waveforms from the spectrograms, a conventional solution is a linear frequency interpolation followed by the Griffin-Lim algorithm, which is
a phase reconstruction method based on the redundancy of the short-time Fourier
transform ~\cite{bayram2012analytic}. However, a straightforward application of this approach only provided us audio with poor audibility: a richness of high-frequency pitches and harmonics made the audio clips harsh and piercing, which makes it impossible to arise ASMR effects on people. These phenomena are mainly caused by the sharp transition at concatenations between different samples. To encounter such effects and \revise{}{further erase the semantic elements, meanwhile }improve the audibility of the synthesized ASMR, we applied moving-average smoothing on the concatenated spectrograms in both frequency and time domains, and added a weak random white noise to the power spectrogram before carrying out the audio reconstruction. These measures significantly smoothed the overall spectrum of the synthesized ASMR audios and enhanced their quality. 
We extracted the same features $\Phi_1$--$\Phi_8$ from the $12$ audios reconstructed from the synthesized samples, as we did with the recorded ASMR audio clips, results are in Tab.~\ref{tab:features_synthesized}.
\begin{table*}[htb]
\centering
\caption{Features extracted from synthesized ASMR audio clips}
\footnotesize
\begin{tabular}{p{2.5cm}<{\raggedright}p{1.3cm}<{\raggedright}p{1.3cm}<{\raggedright}p{1.3cm}<{\raggedright}p{1.3cm}<{\raggedright}p{1.3cm}<{\raggedright}p{1.3cm}<{\raggedright}p{1.3cm}<{\raggedright}p{1.3cm}<{\raggedright}}
	\Xhline{1.2pt}
	\textbf{Audio Clip} & $\mathbf{\Phi_1}$ & $\mathbf{\Phi_2}$ & $\mathbf{\Phi_3}$ & $\mathbf{\Phi_4}$ & $\mathbf{\Phi_5}$ & $\mathbf{\Phi_6}$ & $\mathbf{\Phi_7}$ & $\mathbf{\Phi_8}$\\ 
	\Xhline{1.2pt}
	Synthesized 1 & 4,70E-03 & 1,96E-06 & 0.1390 & 7,52E-05& 7,82E-10 & 0.1250 & 2,39E-03 & 0.9580 \\ 
	Synthesized 2 & 1,09E-02 & 1,20E-05 & 0.1580 & 1,69E-04 & 4,18E-09 & 0.1100 & 2,86E-03 & 0.9500\\ 
	Synthesized 3 & 1,58E-02 & 3,27E-05 & 0.1410 & 1,76E-04 & 9,65E-09 & 0.1400 & 1,40E-03 & 0.9335\\ 
	Synthesized 4 & 5,09E-03 & 2,10E-06 & 0.1310 & 9,30E-05 & 1,18E-09 & 0.1240 & 1,72E-03 & 0.9622\\ 
 \Xhline{1.2pt}
\end{tabular}
\label{tab:features_synthesized}
\end{table*}

\begin{table*}[htb]
\centering
\caption{Overall average scores of recorded and synthesized clips in 5 feeling vectors}
\footnotesize
\begin{tabular}{p{2.0cm}<{\raggedright}p{1.4cm}<{\raggedright}p{1.4cm}<{\raggedright}p{1.4cm}<{\raggedright}p{1.4cm}<{\raggedright}p{2.4cm}}
	\Xhline{1.2pt}
	\textbf{Audio Genre} & Negative & Positive & Attentive & Relaxed & General experience\\ 
	\Xhline{1.2pt}
	Recorded & 1.6242 & 1.5572 & 1.6394 & 2.4230 & 0.6536  \\ 
	Synthesized & 1.5572 & 1.2491 & 1.3876 & 2.0047 & 0.5075 \\ 
 \Xhline{1.2pt}
\end{tabular}
\label{tab:overall_scores}
\end{table*}
\begin{figure*}[!htbp]
\centering
\includegraphics[height=0.208\linewidth]{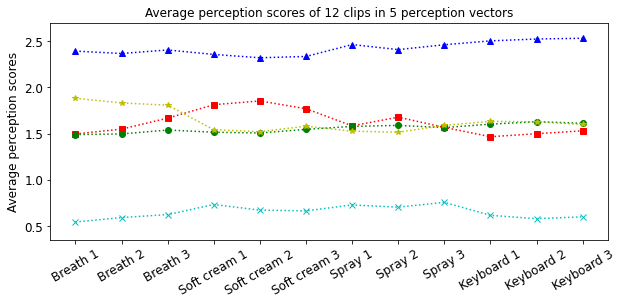}
\includegraphics[height=0.208\linewidth]{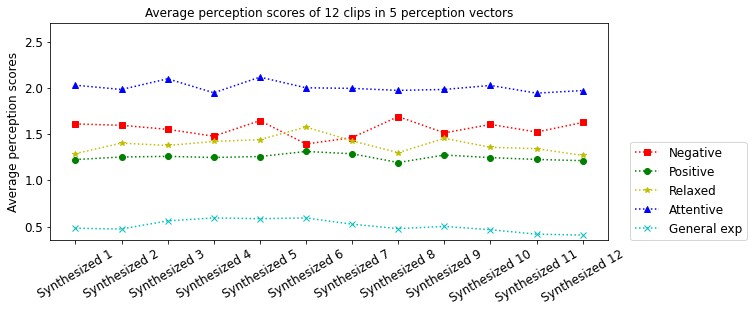}
\caption{Average scores of recorded ASMR clips (left) and synthesized clips (right) in 5 perception vectors}
\label{fig:avescores}
\end{figure*}
\begin{table*}[!htbp]
\centering
\caption{Part of the model fitting results for \emph{Positive} and \emph{Negative} feeling vector}
	\footnotesize
\begin{tabular}{p{2.5cm}<{\raggedright}p{1cm}<{\centering}p{1cm}<{\centering}p{1cm}<{\centering}p{1.3cm}<{\centering}p{1.3cm}<{\centering}p{1cm}<{\centering}p{1cm}<{\centering}p{1cm}<{\centering}}
	\Xhline{1.2pt}
	\textbf{Positive 1st} & $\Phi_1$ & $\Phi_2$ & $\Phi_3$ & $\Phi_4$ & $\Phi_5$ & $\Phi_6$ & $\Phi_7$ & $\Phi_8$ \\ 
	\hline
	Coefficients $\beta_i$& $ 12.158$ & $-18.512$ & $-1.057$ & $-2782.897$ & $-13750.825$ &  $-1.148$ & $-46.432$ & $-0.704$\\
	P-value & $0.005$ & $<0.001$ & $0.056$ & $0.019$ & $0.009$ & $0.167$ & $0.017$	& $0.006$\\ 
	\Xhline{1.2pt}
	\textbf{Negative 1st} & $\Phi_1$ & $\Phi_2$ & $\Phi_3$ & 
	$\Phi_4$ & $\Phi_5$ & $\Phi_6$ & $\Phi_7$ & $\Phi_8$ \\
	\Xhline{1.2pt}
	Coefficients $\beta_i$ & $-11.215$ & $14.540$ & $-0.497$ & $2892.126$ & $4448.740$ &  $2.453$ & $-18.608$ & $0.305$\\
	P-value  & $0.003$ & $0.002$ & $0.308$ & $0.006$ & $0.335$ & $0.001$ & $0.279$ & $0.179$\\
	\Xhline{1.2pt}
	\textbf{Positive 2nd} & $\Phi_1$ & $\Phi_2$ & $\Phi_3$ & 
	$\Phi_4$ & $\Phi_5$ & $\Phi_6$ & $\Phi_7$ & $\Phi_8$ \\
	\Xhline{1.2pt}
	Coefficients $\beta_i$ & $-65.890$ & $5889.119$ & $8.094$ & $3110.414$ & $2555.508$ &  $5.257$ & $-50.452$ & $-3.114$\\
	P-value & $<0.001$ &$ <0.001$ & $<0.001$ &$ <0.001$ & $0.800$ & $0.012$ & $0.004$ & $0.026$\\
	\Xhline{1.2pt}
	\textbf{Negative 2nd} & $\Phi_1$ & $\Phi_2$ & $\Phi_3$ & 
	$\Phi_4$ & $\Phi_5$ & $\Phi_6$ & $\Phi_7$ & $\Phi_8$ \\
	\Xhline{1.2pt}
	Coefficients $\beta_i$& $74.608$ & $-6245.069$ & $-8.235$ & $-3189.256$ &  $-656.987$& $-8.628$ & $62.829$ &  $3.120$\\
	P-value & $<0.001$ & $<0.001$ & $<0.001$ & $<0.001$ & $0.945$  &  $<0.001$ & $<0.001$ & $0.019$ \\
	\Xhline{1.2pt}
\end{tabular}
\label{tab:fitting_res}
\end{table*}
\begin{figure*}[!htbp]
\centering
\includegraphics[width=0.92\linewidth]{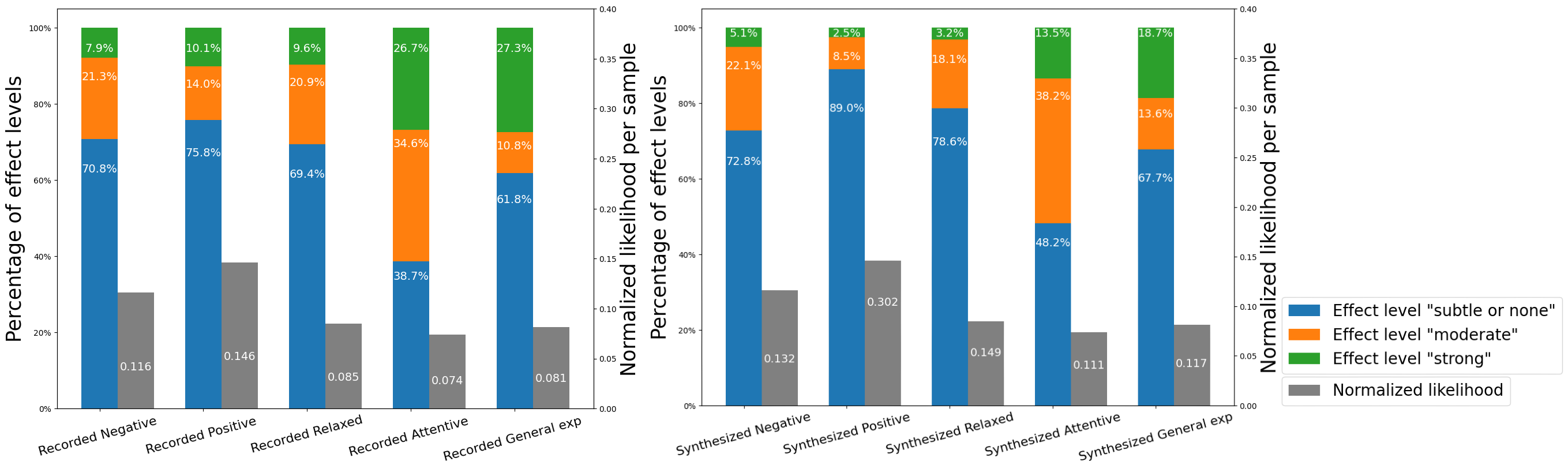}
\caption{Perception scores distribution and likelihood of surveys based on recorded clips (left) and synthesized (right)}
\label{fig:ratio_effect}
\end{figure*}
\clearpage
\twocolumn
\section{BEHAVIORAL STUDY OF PERCEPTIONS}\label{subsec:behavior_study}
To verify the correlation between the ASMR-triggering effect and the acoustic features of an audio, we conducted behavioral experiments on human participants using an online questionnaire survey. U.S.  participants were recruited from Amazon Mechanical Turk (MTurk) for remuneration. After providing informed consent, participants were asked to listen to $13$ audio clips including $1$ white noise and $12$ ASMR audios, each lasting \SI{10}{\second}. The play order was independent and random for every participant to exclude the individual difference in repeated measurements. After listening to each audio clip, the participants reported their emotional state on $10$ dimensions, namely
\begin{enumerate*}[label=\emph{\roman*)}]
\item nervous,
\item irritable
\item upset,
\item delighted,
\item excited,
\item inspired,
\item attentive,
\item concentrating,
\item sleepy, and
\item relaxed,
\end{enumerate*}
which are selected from the widely-used PANAS scales~\cite{WCT1988development}. Participants reported their emotions on 5-point scales ($1$ = ``not at all''; $5$ = ``extremely''). In addition, participants were also asked about their \emph{general experience} whether an ASMR effect had been triggered by the clip on a 3-point scale ($0$ = ``no''; $1$ = ``no ASMR but a precursory ASMR-conductive state''; $2$ 
= ``yes''). For every completed survey, we took the average score of the emotions (\emph{i}--\emph{iii}) as the score for the perception of \emph{negative} feelings, and similarly the averages of (\emph{iv}--\emph{vi}), (\emph{vii}--\emph{viii}), (\emph{ix}--\emph{x}) for \emph{positive}, \emph{attentive}, and \emph{relaxed} feelings, respectively. Thus, a 5D-vector of perception score was obtained from each survey.
The experiment was conducted twice, \revise{the first with $249$ participants and using the recorded ASMR audio clips in Tab.~\ref{tab:features_recorded}, and the second with $251$ participants using the synthesized clips in Tab.~\ref{tab:features_synthesized}.}{the first one was carried out in February 2022 using the recorded ASMR audio clips in Tab.~\ref{tab:features_recorded} , with $249$ participants and $3237$ observations included. The second one was in August 2022 using the synthesized clips in Tab.~\ref{tab:features_synthesized}, with $251$ participants and $3263$ observation included.} After each experiment, we applied regression analysis on the collected set of scores w.r.t. $\Phi_1$--$\Phi_8$, using the linear mixed-effect model (LMM):
\begin{equation}
y=\beta_0+\sum\limits_{i=1}^8\beta_i\Phi_i+\sum\limits_{i=0}^8b_iz_i+\epsilon,\label{eq:lmm}
\end{equation}
where $y$ can be used to fit every dimension of the perception score vector. $\epsilon$ is the model error, $\beta_0$ the constant offset, $\beta_1$--$\beta_8$ the linear coefficients regarding features $\Phi_1$--$\Phi_8$, respectively. The random variables $z_1$--$z_8$ weighted by coefficients $b_1$--$b_8$ are introduced to capture the random effect caused by repeated measures, which is suggested as a standard approach in cognitive neuroscience and experimental psychology~\cite{Magezi2015linear}. Fig.\ref{fig:avescores} presents the survey results of average scores for 12 recorded audio clips and 12 synthesized clips, as rated by participants in 5 perception vectors. Tab.\ref{tab:overall_scores} presents the overall scores of recorded and synthesized clips. Survey results in Fig. \ref{fig:avescores} show significant score differences between groups that presented specific scenarios. Breath clips had the highest relaxed perception scores (1.8420), while clicking keyboard clips had the highest attentive perception scores (2.5201). Synthesized clips do not manifest the same correlation as recorded clips, possibly due to not associating with specific scenarios.
A summary of the survey scores and the likelihood of regression is shown in Fig.\ref{fig:ratio_effect}, in which we specify the average score range $0-2$ points as effect level \emph{subtle or none}, score range $2-3$ as effect level \emph{moderate}, and $3-5$ as effect level \emph{strong}. Similarly, for vector General experience, we specify score $0$ as effect level \emph{subtle or none}, score $1$ as effect level \emph{moderate}, and score $2$ as effect level \emph{strong}. Compared with the recorded clips, \revise{our synthesized audio clips  are performing similarly in triggering emotional perceptions and ASMR experiences, with a slightly enhanced suppression of negative emotions, but the weaker effect in other perspectives}{our synthesized audio clips are performing similarly in triggering \emph{moderate} ASMR effect (similar ratio over all perception vectors in Fig.~\ref{fig:ratio_effect}) but weaker in triggering \emph{strong} ASMR effect (decreased ratio over all perception vectors in Fig.~\ref{fig:ratio_effect}), which not only indicates that ASMR effect can rise from meaningless acoustic patterns but also confirms that semantic elements play a significant role in bringing audience strong ASMR experience}. Even though the likelihood indexes of second survey are enhanced but still not strong enough to prove a linear relationship between perception scores and extracted features, which is not surprising due to the high non-linearity of human neural systems and cognition processes. Tab.\ref{tab:fitting_res} shows the fitting results of our model, which confirm the validity of our approach. The coefficients $\beta_1 - \beta_8$ for the composite feeling vectors \emph{positive} and \emph{negative} mostly exhibit opposite polarity. Although the coefficients $\beta_1 - \beta_8$ for the two surveys were not consistent, they still displayed opposite polarity. This suggests that the human brain may react differently to identifiable and non-identifiable sounds. Our results reveal a significant correlation between the ASMR triggering effect and the acoustic features $\Phi_1-\Phi_8$, with our generally very low p-values as evidence.
\section{CONCLUSION AND OUTLOOKS}
In this paper, we have studied the ASMR phenomenon from an acoustic perspective, and successfully affirmed a significant correlation between the ASMR triggering effect and the cyclic spectral characteristics of audio. Crucially, we have implemented a DCGAN-based system, which synthesizes random artificial audio clips that effectively trigger ASMR.

For future research, two approaches can be interesting. From the signal processing and AI perspective, a novel application-specific design of DCGAN that directly generates audio waveform without relying on spectrogram images may further improve the quality and ASMR triggering effect of the synthesized audios. From the psychological perspective, it is important to note the limitations of modelling human behaviors. Replacing behavioral study with neuroimaging techniques may help us better observe the ASMR effect and understand the psychological mechanism behind it.

\bibliographystyle{IEEEtran}
\bibliography{IEEEabrv,mylib}


\end{document}